\documentclass[aps,prd,twocolumn,amsmath,amssymb,nofootinbib,superscriptaddress,floatfix,10pt]{revtex4-1}

\usepackage{dsfont}
\usepackage{graphicx}
\usepackage{hyperref}
\usepackage[utf8]{inputenc}
\usepackage{makecell}
\usepackage{microtype}
\usepackage{placeins}
\usepackage{slashed}
\usepackage[caption=false]{subfig}
\usepackage{xcolor}

\newcommand{\cc}{\langle\psibar\psi\rangle}
\newcommand{\D}{\mathcal{D}}
\newcommand{\Dov}{D_\mathrm{ov}}
\newcommand{\Dw}{D_W}
\newcommand{\g}{\gamma_5}
\newcommand{\id}{\mathds{1}}
\newcommand{\ii}{\mathrm{i}}
\newcommand{\muc}{\mu_{\mathrm{c}}}
\newcommand{\Nf}{N_{\mathrm{f}}}
\newcommand{\Ns}{N_{\mathrm{s}}}
\newcommand{\Nt}{N_{\mathrm{t}}}
\renewcommand{\O}{\mathcal{O}}

\newcommand{\sigex}{\langle\vert\bar{\sigma}\vert\rangle}
\newcommand{\psibar}{\bar{\psi}}
\newcommand{\Veff}{V_\mathrm{eff}}
\newcommand{\Z}{\mathbb{Z}}

\newcommand{\sref}[1]{Sec.~\ref{#1}}
\newcommand{\aref}[1]{App.~\ref{#1}}
\newcommand{\tref}[1]{Tab.~\ref{#1}}
\newcommand{\fref}[1]{Fig.~\ref{#1}}

\renewcommand{\eqref}[1]{Eq.~(\ref{#1})}
\renewcommand{\L}{\mathcal{L}}

\DeclareMathOperator{\cov}{cov}

\graphicspath{{./figures/}}

\begin{document}

	\title{The magnetized (2+1)-dimensional Gross-Neveu model at finite density}
	
	\author{Julian J. Lenz}
	\email{j.j.lenz@swansea.ac.uk}
	\affiliation{Theoretisch-Physikalisches Institut, Friedrich-Schiller-Universität Jena,  D-07743 
	Jena, Germany}
	\affiliation{Swansea Academy of Advanced Computing, Swansea University, Fabian Way, SA1 8EN, 
	Swansea, Wales, UK}
	\author{Michael Mandl}
	\email{michael.mandl@uni-jena.de}
	\affiliation{Theoretisch-Physikalisches Institut, Friedrich-Schiller-Universität Jena,  D-07743 
	Jena, Germany}
	\author{Andreas Wipf}
	\email{wipf@tpi.uni-jena.de}		
	\affiliation{Theoretisch-Physikalisches Institut, Friedrich-Schiller-Universität Jena,  D-07743 
	Jena, Germany}
	
\begin{abstract}	
	We perform a lattice study of the ($2+1$)-dimensional Gross-Neveu model in a background 
	magnetic field $B$ and at non-zero chemical potential $\mu$. The complex-action problem arising 
	in our simulations using overlap fermions is under control. For $B=0$ we observe a first-order 
	phase transition in $\mu$ even at non-vanishing temperatures. Our main finding, however, is that 
	the rich phase structure found in the limit of infinite flavor number $\Nf$ 
	is washed out by 
	the fluctuations present at $\Nf=1$. We find no evidence for inverse magnetic catalysis, i.e., 
	the decrease of the order parameter of chiral symmetry breaking with $B$ for $\mu$ close to the 
	chiral phase transition. Instead, the magnetic field tends to enhance the breakdown of chiral 
	symmetry for all values of $\mu$ below the transition. Moreover, we find no trace of spatial 
	inhomogeneities in the order parameter. We briefly comment on the potential relevance of our 
	results for QCD.
\end{abstract}		
	\maketitle

	\section{Introduction}\label{sec:introduction}
	The study of Quantum Chromodynamics (QCD) at finite baryon density is a highly non-trivial endeavor 
due to the complex-action problem, which prevents the use of lattice simulations based on 
importance sampling \cite{TW05}, the most reliable 
\emph{ab initio} tool for the non-perturbative study of 
strongly interacting matter.  With lattice QCD no longer at one's disposal in a parameter regime 
that is, e.g., relevant for the physics of compact stellar objects like neutron stars, an 
alternative is much needed. While considerable effort is put into finding methods that circumvent 
the complex-action problem, another approach entirely is the study of low-energy 
effective theories, which reproduce QCD phenomenology within their range of validity. 

Prominent examples of such effective field theories are
those based on chiral perturbation theory
\cite{Sch03} and the
four-Fermi theories (4FTs). The latter arise in the low-energy limit of QCD \cite{Kon10} and are 
capable of capturing a number of essential features of QCD, in particular,
chiral symmetry and its  spontaneous breakdown. 
There are examples of 4FTs that are amenable 
to lattice studies at finite density, since they do not suffer from a complex-action problem due to 
their rather simple structure -- see, e.g., \cite{HKK93}. In fact, a great part of our current 
understanding of finite-density QCD stems from the investigation of 4FTs \cite{VW91,Kle92,Str03}.

One particularly interesting question is how the structure of strongly interacting matter changes 
under the influence of background magnetic fields \cite{Sho13r,MS15r,Ant16r}. This is due to the 
fact that magnetic fields of the order of the QCD scale are generated in non-central heavy-ion 
collisions \cite{Tuc13}, are present in the cores of magnetars, \cite{FIK10} and were likely
produced during the electroweak phase transition \cite{Vac91}. However, because of the 
aforementioned limitations of lattice simulations, magnetized systems at finite baryon density are 
still quite elusive.

To this end, we perform in this work a lattice study of the Gross-Neveu (GN) model \cite{GN74}, the 
simplest 4FT, in $(2+1)$ space-time dimensions. Extending our previous work \cite{LMW23}, which was 
concerned with the magnetized GN model at zero density but finite temperature, we here work at low 
temperature but non-zero chemical potential. In \cite{LMW23}, it became clear that this simple model 
fails to correctly describe the phenomenology of magnetized QCD \cite{BBE12_2,BBE12} both in and 
beyond the mean-field limit. However, we also emphasized its role as a starting point for the 
description of QCD in background magnetic fields by means of beyond-mean-field effective models.

It shall be one of our goals to shed light on the question of how much of the rich phase structure 
the model exhibits in the mean-field limit \cite{VKM96} persists when quantum fluctuations are 
taken into account. Work in this direction has already been done using the optimized perturbation 
theory (OPT) technique \cite{KPR13}, but, to the best of our knowledge, there exist no \emph{ab 
initio} lattice simulations in the literature that are concerned with that question. Furthermore, 
we investigate whether the magnetic field induces spatial inhomogeneities at finite density 
as it likely does at very strong fields
in $3+1$ dimensions \cite{BDK10,FZK10,TNK15,BC16}. Lastly, as a long-term goal, we aim at 
understanding properly to what extent our findings are of relevance for QCD.

We provide access to our simulation data online \cite{data} in order to ensure the 
reproducibility of our results in accordance with the FAIR\footnote{For a recent update on the 
status of Open Science within the lattice community, see \cite{ABLP22}.} guiding principles 
\cite{FAIR16}. Moreover, our data analysis scripts can be found in \cite{code}.

The outline of this work is as follows. In \sref{sec:analytical}, we introduce the GN model and 
discuss how chiral symmetry and its spontaneous breakdown are affected by a chemical potential and 
an external magnetic field in the mean-field limit. In particular, we discuss the complicated phase 
structure arising due to fermionic Landau levels. \sref{sec:lattice} outlines our lattice formalism 
using overlap fermions, putting a particular emphasis on the complex-action problem present in our 
simulations and how it is avoided. We then present our simulation results obtained at finite 
density and magnetic field in \sref{sec:results} before discussing their relevance in 
\sref{sec:discussion}. A large part of our formalism and notation was introduced in \cite{LMW23},
and we shall refer to that work on various occasions for brevity.
	
	\section{Analytical results}\label{sec:analytical}
	The GN model at finite density (determined by the chemical potential $\mu$) and magnetic field 
(described by the vector potential $A_\mu$) in the chiral limit is defined by the Lagrangian
\begin{equation}\label{eq:gn_model}
	\L_\sigma = \ii\psibar\left(\slashed{\partial} + 
	\ii e\slashed{A} + 
	\sigma + 
	\mu\gamma_0\right)\psi + 
	\frac{\Nf}{2g^2}\sigma^2\;,
\end{equation}
where $e$ is the elementary electric charge, $\Nf$ denotes the number of fermion flavors (the sum
over flavors is implicit in (\ref{eq:gn_model})), and $g^2$ denotes the four-Fermi coupling 
constant. To arrive at \eqref{eq:gn_model}, we have performed the usual Hubbard-Stratonovich 
transformation introducing the auxiliary scalar field $\sigma$ in exchange for the 
$(\psibar\psi)^2$ term.

\begin{figure*}[t]
	\hspace{0.55cm}
	\captionsetup[subfigure]{margin={1.1cm,0cm}}
	\subfloat{
		\includegraphics[width=0.40\linewidth]{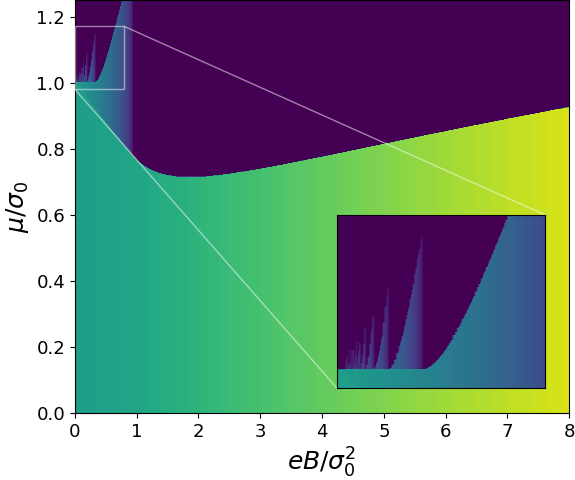}}
	\hspace{0.5cm}
	\captionsetup[subfigure]{margin={-1.5cm,0cm}}
	\subfloat{
		\includegraphics[width=0.445\linewidth]{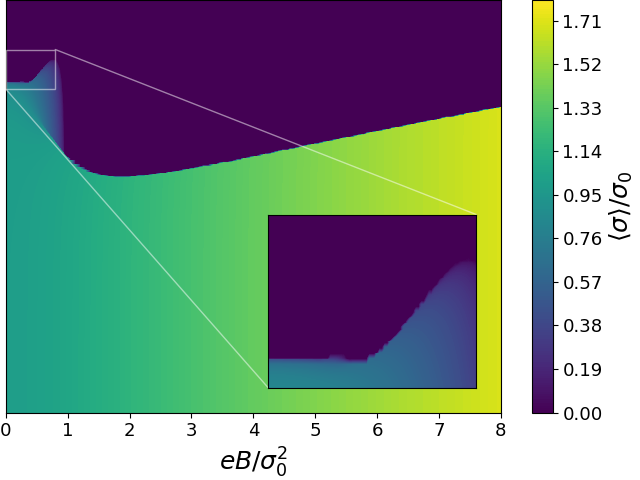}}
	\caption{$(B, \mu$) phase diagram of the $(2+1)$-dimensional GN model in 
	the mean-field limit at $T=0$ (left) and $T=0.1\sigma_0$ (right). The insets show enlarged regions where multiple phase transitions 
	occur.}
	\label{fig:pd_large_N}
\end{figure*}

In this work, we consider a three-dimensional Euclidean space-time and work with four-component 
spinors, which allows for the definition of a matrix $\g$, anti-commuting with all other gamma 
matrices. The model then has a $\Z_2$ chiral symmetry, \footnote{Strictly speaking, there is no chiral symmetry in odd dimensions. Here, it refers to the symmetry in the reducible representation inherited from chiral symmetry in four dimensions.} being invariant under the simultaneous 
transformations
\begin{equation}
	\psi \to \g\psi\;, \quad \psibar \to -\psibar\g\;, \quad \sigma \to -\sigma\;.
\end{equation}
This chiral symmetry may be spontaneously broken by the formation of a chiral condensate $\cc$, 
which can be shown to be related to the expectation value of $\sigma$ by means of a Dyson-Schwinger 
equation:
\begin{equation}\label{eq:ward_identity}
	\cc = \frac{\ii\Nf}{g^2}\langle\sigma\rangle\;.
\end{equation}

In \cite{LMW23} we presented a computation of the effective potential $\Veff$ of the GN model in 
$2+1$ dimensions in the limit $\Nf\to\infty$, where the mean-field approximation becomes exact. 
Assuming translational invariance in space and time, $\sigma(x)=\sigma=const.$, and that the 
magnetic field lies perpendicular to the spatial plane and has a magnitude $B$ such that, without 
loss of generality, $eB>0$, one finds (see also \cite{GMS95})
\begin{align}\label{eq:effective_potential}
	\begin{aligned}
		\Veff(\sigma)
%		\frac{\Veff(\sigma)}{L^2} 
		= -\frac{\sigma^2}{2\pi}\sigma_0 
		 - \frac{\sqrt{2}}{\pi}(eB)^{3/2}\zeta_H\left(-\frac{1}{2}, \frac{\sigma^2}{2eB}\right)
	     + \frac{\vert\sigma\vert eB}{2\pi} \\
	    - \frac{eB}{2\pi\beta}\sum_{l=0}^{\infty}d_l
	    	\left[\ln\left(1+e^{-\beta\left(\sqrt{\sigma^2+2eBl}+\mu\right)}\right) + 
	    	(\mu\leftrightarrow-\mu)\right]\;,
	\end{aligned}
\end{align}
where $\zeta_H$ denotes the Hurwitz zeta function and $\beta=1/T$ 
is the inverse temperature. The sum runs over the fermionic Landau levels, labeled by the 
index $l$, and $d_l=2-\delta_{l0}$ takes into account that the degeneracy of the lowest Landau 
level (LLL) is only half of that of the higher ones. 

Because of (\ref{eq:ward_identity}), the chiral condensate in the large\,-$\Nf$ limit is proportional to 
the position of the global minimum of $\Veff(\sigma)$, i.e., to the solution $\langle\sigma\rangle$ 
of the gap equation
\begin{equation}
	\Veff'(\sigma)\big\vert_{\sigma=\langle\sigma\rangle} = 0\;.
\end{equation} 
In the following, we denote by $\sigma_0$ the value of $\langle\sigma\rangle$ at zero temperature, 
chemical potential, and magnetic field. We are interested in the phase structure of the model at 
finite chemical potential and vanishing to low temperature. To this end, we have performed a 
minimization of $\Veff$ in the $(B,\mu)$ plane, and we show the $T=0$ phase diagram in 
\fref{fig:pd_large_N} (left).

A striking feature of the $(B,\mu)$ phase structure at zero temperature is the cascade of 
first-order\footnote{We remark that for $B=0$ the (single) phase transition is of second order 
everywhere but at the point $(T=0, \mu=\sigma_0$), where it becomes degenerate.} phase transitions 
in $\mu$ for small $eB$. The physical origin of these multiple phase transitions lies in the 
discreteness of Landau levels. As long as $B$ is small, the Landau levels are closely spaced, such 
that for increasing chemical potential the Fermi energy crosses them successively, resulting in the 
possibility for the order parameter to jump discontinuously for every such crossing. When the 
magnetic field is strong enough, however, the energy difference between Landau levels grows too 
large and only the LLL remains occupied, such that only the chiral phase transition (i.e., the 
transition from $\langle\sigma\rangle\neq0$ to $\langle\sigma\rangle=0$), but no intermediate 
transition, is seen.

As can be seen in \fref{fig:pd_large_N} (right), thermal fluctuations present at $T\neq0$ wash 
out the pattern of multiple phase transitions. This can be understood by recalling that at finite 
temperature the underlying Fermi-Dirac distribution is no longer a step function but becomes 
smoother, which, in turn, results in a smoother behavior of the order parameter as the Landau levels 
are crossed. Still, even at $T/\sigma_0=0.1$, an intermediate phase can be found for small $eB$ and 
large $\mu$. We also mention that the critical chemical potential $\muc$ of the chiral phase 
transition shows a non-monotonic behavior in $B$ as long as the latter is not too strong, while it 
grows monotonically for larger $eB$.

Moreover, one observes that generically the phase diagram is roughly divided into two regions: the 
large \emph{magnetic catalysis} region, where the order parameter increases with the magnetic field, 
and the smaller \emph{inverse magnetic catalysis} region, where it decreases with $B$. We emphasize 
the stark contrast to the situation at zero density studied in \cite{LMW23}, where only magnetic 
catalysis is present for all magnetic field strengths and temperatures.

A physical explanation for magnetic catalysis is provided in \cite{GMS94} by the effective reduction 
of the number of space-time dimensions due to the presence of the magnetic field, which causes 
infrared divergences to which the system responds via the formation of a mass gap. The inverse 
magnetic catalysis found for weak magnetic fields and large chemical potentials, on the other hand, 
was explained in \cite{PRS11,PRS13} to be caused by a competition between the energy gain due to the 
formation of a chiral condensate (which increases with $B$) and the energy cost of overcoming the 
imbalance between fermions and anti-fermions at finite $\mu$ (which increases with both $B$ and 
$\mu$). 

Note that in the context of finite-temperature QCD the expression ``inverse magnetic catalysis'' 
commonly refers to the decrease of the chiral cross-over temperature with $B$, accompanied by a 
non-monotonic $B$-dependence of the chiral condensate \cite{BBE12,BBE12_2}. One should, however, be 
careful when comparing the situation in QCD to the one considered here, since their physical 
origins appear to be quite different.

Finally, we mention that the lattice study \cite{KS01} provided evidence for the existence of a 
tri-critical point in the $(T,\mu)$ plane at $B=0$, accompanied by a first-order transition line 
for non-vanishing temperatures, in contradiction to the known mean-field results. While in 
analytical beyond-large\,-$\Nf$ studies such as the OPT calculations \cite{KPR07,KPR07_2} a similar 
result was found, we argue that one may encounter first-order transitions at the mean-field level as 
well, provided that one studies the theory on a finite spatial volume. The reasoning is as follows: 
On a finite volume, the allowed momenta and, thus, the one-particle energies are discrete, which can 
give rise to discontinuous phase transitions in the same way as the Landau quantization. In a way, 
the $B=0$ theory in a finite volume is thus reminiscent of the $B\neq0$ theory.

More concretely, the GN effective potential for vanishing magnetic field on a finite spatial volume 
$L^2$, such that the space-time volume reads $V=\beta L^2$, is given by
\begin{widetext}
	\begin{align}\label{eq:gn_effective_potential_finite_volume}
		\begin{aligned}
			\Veff(\sigma)\Big\vert_{B=0}=
			%\frac{\Veff(\sigma)}{L^2} \bigg\vert_{B=0} = \\
				- \frac{\sigma^2}{2\pi}\sigma_0 + 
				\frac{\vert\sigma\vert^3}{3\pi} 
				+ \frac{\sigma}{\pi L^2}
				{\sum_{\mathbf{n} }}^{\prime}
				%\sum_{(n_1, n_2)}{\vphantom{\sum}}'
					e^{-L\sigma\vert\mathbf{n}\vert}
					\frac{1}{\mathbf{n}^2}\left(1+\frac{1}{L\sigma\vert\mathbf{n}\vert}\right)
		   		 -\frac{2}{V}\sum_{\mathbf{p}}
					\left[\ln\left(1+e^{-\beta(\sqrt{\sigma^2+\mathbf{p}^2}+\mu)}\right) +
				      (\mu\leftrightarrow-\mu)\right]\;,
		\end{aligned}
	\end{align}
\end{widetext}
where $\mathbf{n}=(n_1,n_2)\in\Z^2$ and the prime on the first sum indicates omission of 
the summand where
$n_1=n_2=0$, while the second sum runs over spatial momenta $\mathbf{p}=\frac{2\pi}{L}\mathbf{n}$. 
The derivation is similar as in the $B\neq0$ case, with the sum over Landau 
levels being replaced by momentum sums; see, e.g., \cite{Wipf:2021mns}.

The first sum in (\ref{eq:gn_effective_potential_finite_volume}) represents the
finite-size 
corrections to $\Veff$, while the last term is the analog of the last term in 
(\ref{eq:effective_potential}) on a finite volume and for vanishing magnetic field, and thus -- 
notice their similarity -- may also give rise to discontinuities at low temperatures. We remark 
that $\sigma_0$ in (\ref{eq:gn_effective_potential_finite_volume}) refers to the value of the 
condensate for vanishing $B$, $T$, and $\mu$ in the infinite-volume limit, as above, and not its
finite-volume counterpart.

\begin{figure}[t]
	\centering
	\includegraphics[width=0.95\linewidth]{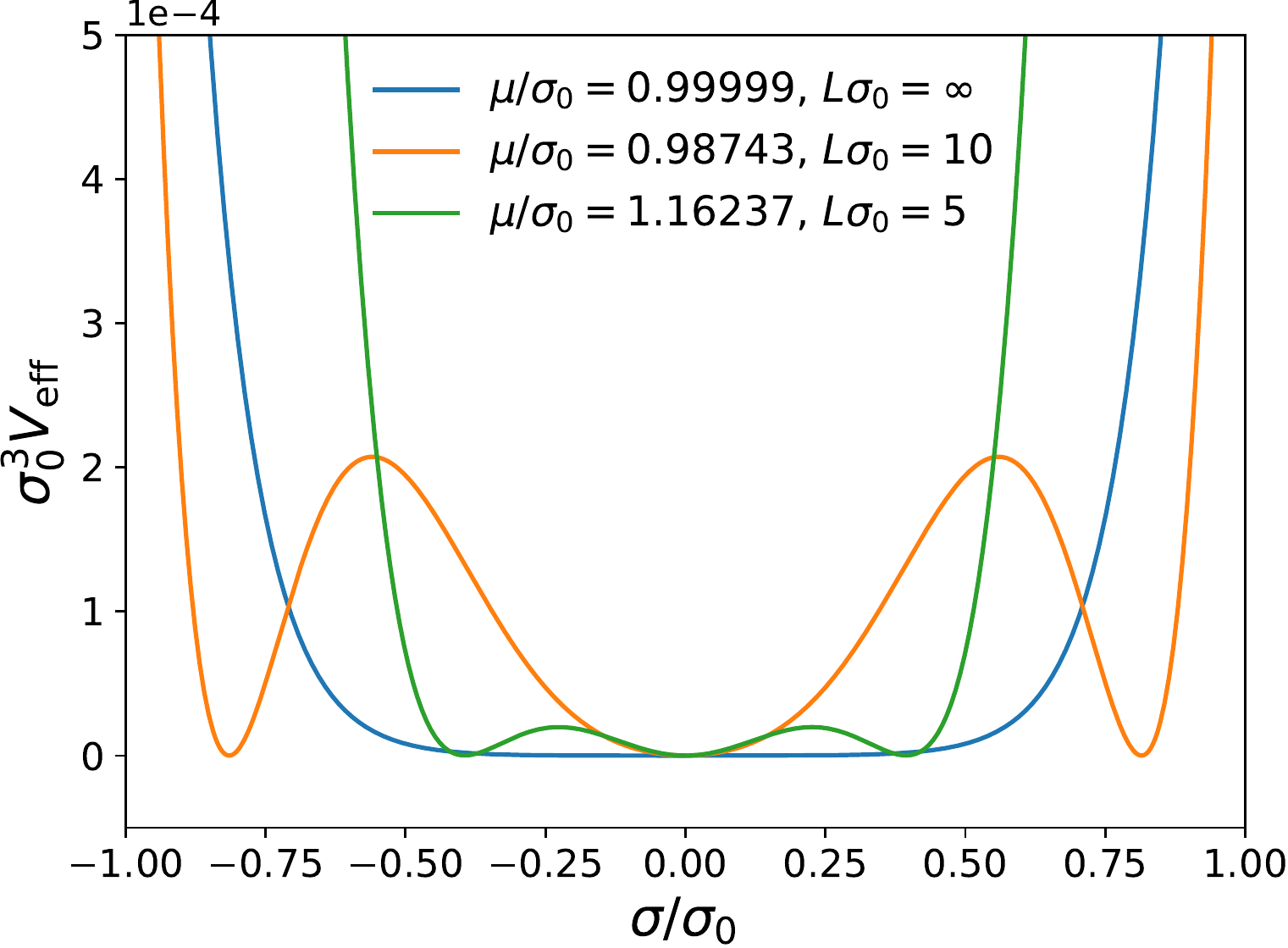}
	\caption{$\Veff$ as a function of $\sigma$ for $T/\sigma_0=0.1$ at the chiral phase 
	transition, $\mu=\muc(T)$, and for different $L$.}
	\label{fig:veff_large_N}
\end{figure}

We show in \fref{fig:veff_large_N} a comparison between the effective potential at the 
chiral phase transition for both finite and infinite $L$ 
(always assumed equal in both directions) at a low non-vanishing temperature. 
We see that on the finite volumes $\Veff$ exhibits three degenerate global minima: one at 
$\sigma=0$ and two at non-trivial values of $\sigma$ related to one another via a chiral 
transformation. The minima are separated by potential barriers, which is indicative of a 
first-order phase transition, since it implies the coexistence of two phases. For $L=\infty$, on the 
other hand, the non-trivial minima turn into the trivial one in a smooth way when increasing $\mu$, 
which rather hints at a second-order transition. A more detailed analysis on finite-volume effects 
will be the subject of a forthcoming publication.

Since our lattice simulations are all performed on finite 
volumes, they could conceivably reveal a first-order transition as well, leading us to investigate 
this question in more detail below. We remark, however, that a weak non-vanishing magnetic field on 
finite volumes and at non-zero temperatures might, in fact, drive the system back to a second-order 
phase transition, which then becomes first order again only for strong enough
$B$. Also, there are examples where first-order phase transitions found under
the assumption of homogeneity were later understood to be of second order after
lifting the latter constraint \cite{TU03}. In \cite{LPW20}, we developed
the technology to investigate this on the lattice.

Finally, we emphasize the non-monotonic behavior of the critical chemical potential with $L$ in 
\fref{fig:veff_large_N}, which is reminiscent of the $B$-dependence of $\muc$ for a sufficiently weak 
magnetic field. From these observations, one clearly realizes that the interplay between non-zero 
magnetic field, temperature, and chemical potential has a highly non-trivial influence on the order 
parameter even in the infinite-volume limit and becomes even more involved once $L<\infty$ enters 
as an additional control parameter. We also mention the possibility of introducing a tilt in the magnetic field, resulting in a non-trivial phase structure even at zero chemical potential \cite{KZ13,ZKZ16}.

	\section{Numerical setup}\label{sec:lattice}
	\subsection{Simulations with overlap fermions}
We perform lattice simulations of the $(2+1)$-dimensional GN model using one 
reducible four-component flavor, $\Nf=1$, of 
Neuberger's overlap fermions \cite{Neu98}. We couple the chemical potential $\mu$ in a way 
suggested by Gavai and Sharma \cite{GS12}, such that our full lattice Dirac operator reads
\begin{equation}
	D = (\sigma+\mu\gamma_0)\left(\id-\frac{a}{2}\Dov\right) + \Dov\;,
\end{equation}
where $\Dov$ is the massless overlap operator,
\begin{equation}
	\Dov = \frac{1}{a}\left(\id + \Dw(-1)/\sqrt{\Dw^\dagger(-1)\Dw(-1)}\right)\;,
\end{equation}
$a$ denotes the lattice constant, and $\Dw(am)$ is the standard Wilson operator with mass $m$. $\Dw$ 
also contains the $U(1)$ link variables encoding the magnetic field. For more details on the 
discretization, including a thorough discussion on how the magnetic field is implemented in our 
simulations as well as a number of numerical tests, we refer to \cite{LMW23}. 

There, we also argue that the lattice chiral condensate \cite{Cha99} is related to the expectation 
value of $\sigma$ via
\begin{equation}
	\left\langle\psibar\left(\id-\frac{a}{2}\Dov\right)\psi\right\rangle = 
	-\frac{\Nf}{g^2}\langle\sigma\rangle\;,
\end{equation}
in analogy to the continuum expression (\ref{eq:ward_identity}). As in \cite{LMW23}, we use the 
observable $\sigex$, where $\bar{\sigma}$ denotes the space-time average of $\sigma$, as an order 
parameter for chiral symmetry breaking and we perform the scale setting at $\mu=B=0$ and at low 
temperatures. A detailed list of the parameter values used in our simulations can be found in 
\aref{app:parameters}.

\subsection{The complex-action problem}
In \cite{LMW23}, we showed that our lattice action is real for arbitrary $B$ at vanishing $\mu$ and  
the same holds true in the case $B=0$ and $\mu\neq0$. In this work, we are, however, concerned with  
$B$ and $\mu$ both being non-zero, such that we have to expect a complex-action problem in general.
However, we found numerically that the ensuing complex-action problem is, in
fact, mild -- in particular, with respect to the estimation of the chiral condensate. 
The fluctuating phase then gives rise only to a systematic uncertainty, which we estimate on 
exemplary ensembles. Most importantly, we demonstrate in the following that it
is negligible compared to the statistical uncertainties. 

\begin{figure*}[t]
	\captionsetup[subfigure]{margin={0.7cm,0cm}}
	\subfloat{
		\includegraphics[width=0.35\linewidth]{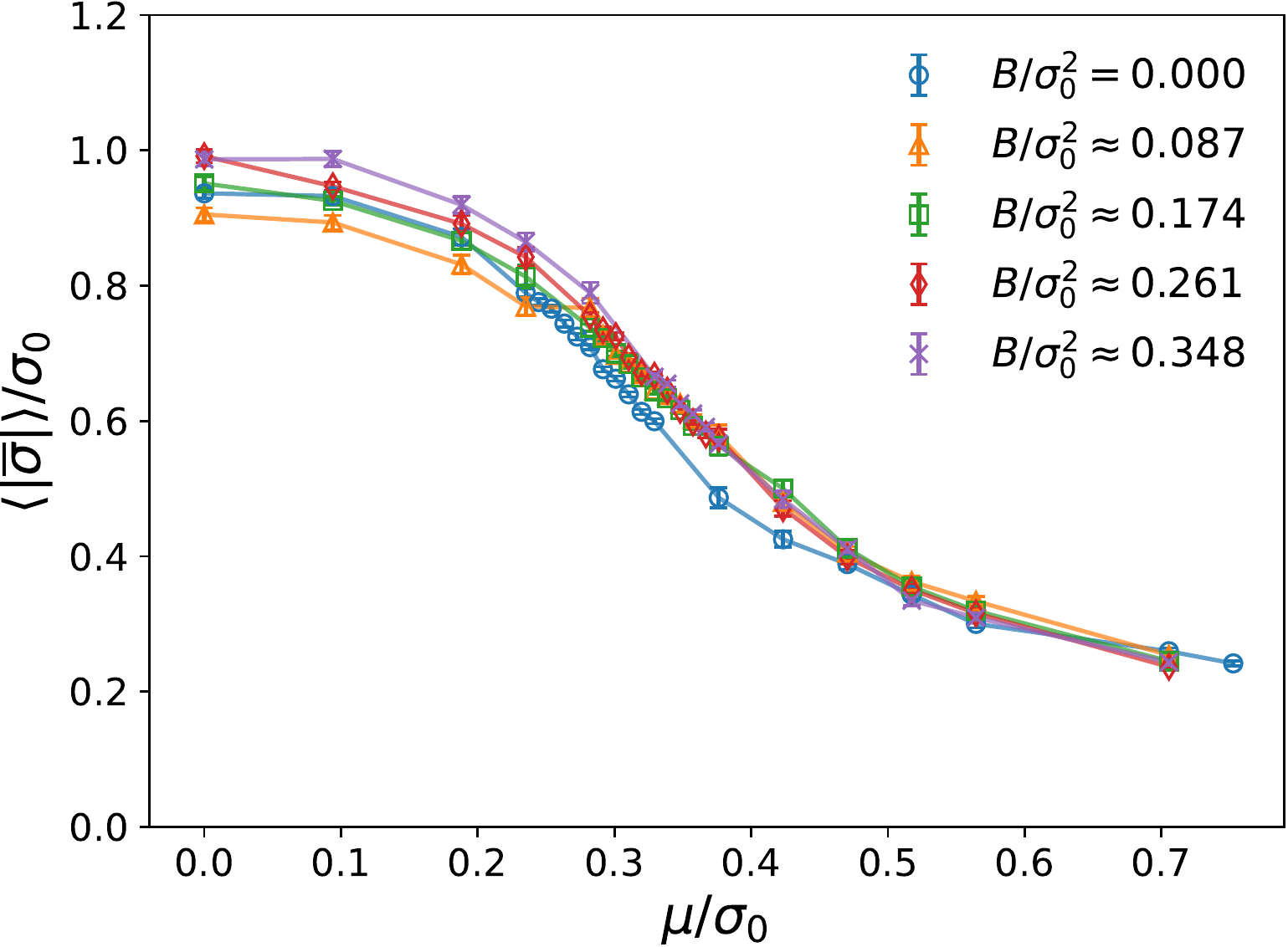}}
	\captionsetup[subfigure]{margin={-0.1cm,0cm}}
	\subfloat{
		\includegraphics[width=0.3105\linewidth]{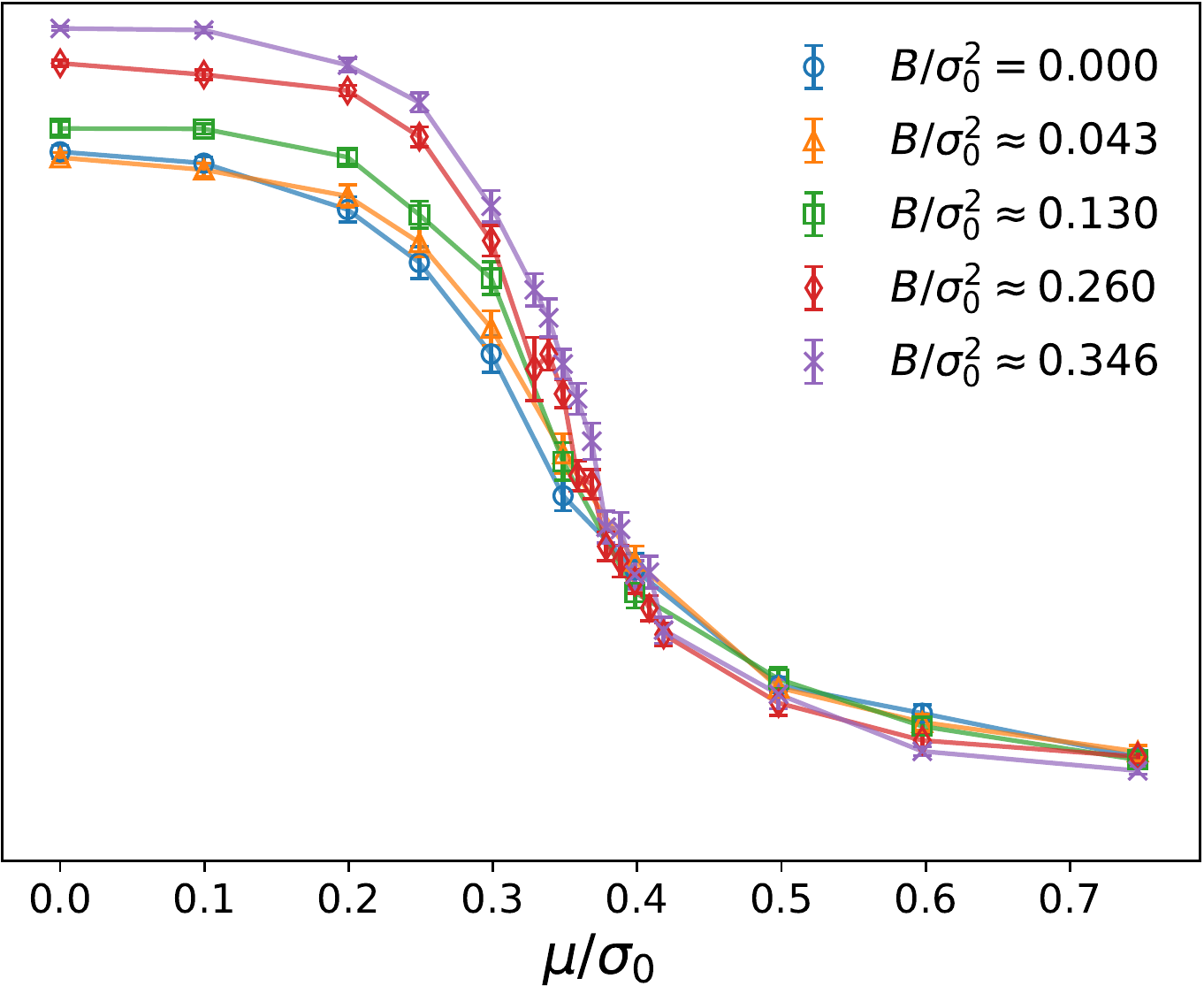}}
	\subfloat{
		\includegraphics[width=0.3105\linewidth]{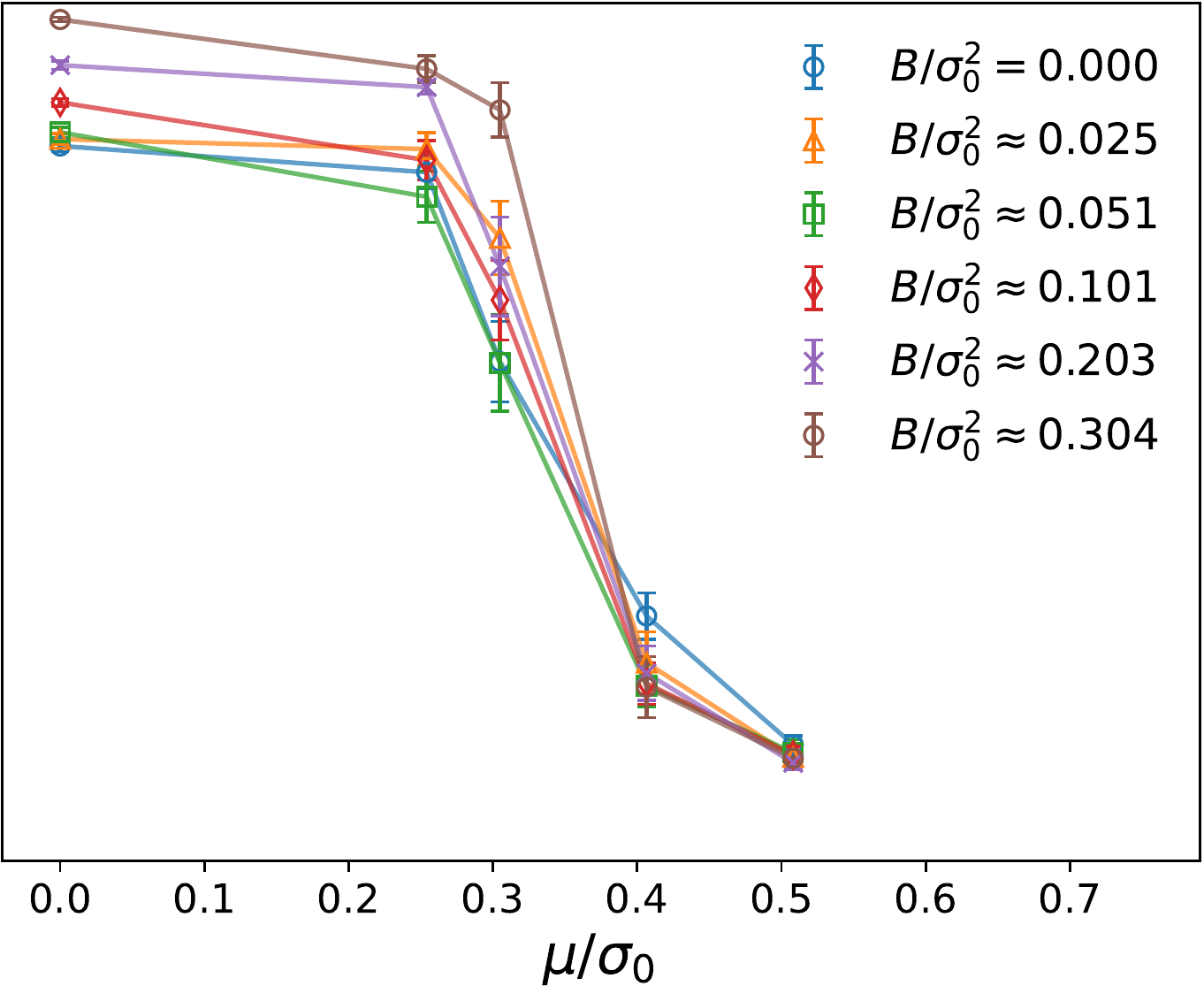}}
	\caption{Order parameter $\sigex$ as a function of $\mu$ for different magnetic field strengths 
	in an infinite-volume extrapolation at fixed lattice spacing. (left) $\Ns=8$, $a\sigma_0\approx1.063$. (center) $\Ns=12$, $a\sigma_0\approx1.004$. (right) $\Ns=16$, $a\sigma_0\approx0.984$. We consider only cubic lattices 
	where $\Ns=\Nt$\;.}
	\label{fig:cc_vs_mu_infVol}
\end{figure*}

For the expectation value of an observable $\O$, the standard re-weighting approach
\begin{equation}\label{eq:reweighting}
	\langle\O\rangle = \frac{\int\D\sigma e^{-S_R - \ii S_I}\O}{\int\D\sigma e^{-S_R - \ii S_I}} = 
	\frac{\langle e^{-\ii S_I}\O\rangle_R}{\langle e^{-\ii S_I}\rangle_R}
\end{equation}
provides an exact representation with stochastic interpretation in the
presence of a complex action $S=S_R + \ii S_I$ with $S_R, S_I\in \mathbb{R}$. Here,
$\langle\,\cdot\,\rangle_R$ denotes the expectation value with respect to the 
probability distribution $e^{-S_R}/Z_R$ for an appropriate normalization $Z_R$.
As is well known, this expression does not
solve the complex-action problem because the numerical determination of the quotient
on the right-hand side of (\ref{eq:reweighting}), in general, requires an exponential 
amount of computational resources in
the thermodynamic limit \cite{GL16}. However, this statement is
concerned with only the asymptotic behavior, and, depending on the observable in
question, the desired parameter regime might still be reachable at a reasonable
numerical cost.

More precisely, if the covariance between $\O$ and $e^{\ii S_I}$ is negligible
compared to the average phase $\langle e^{-\ii S_I}\rangle_R$, the latter
approximately drops out of the expectation value:
\begin{equation}\label{eq:expectation_value_factorization}
	\langle\O\rangle = \langle\O\rangle_R + 
	\frac{\cov_R\left(e^{-\ii S_I},\O\right)}{\langle e^{-\ii S_I}\rangle_R}
  \approx \langle\O\rangle_R\;,
\end{equation}
where the covariance between two random variables $X$ and $Y$ is defined as
\begin{equation}\label{eq:covariance}
	\cov_R\left(X,Y\right) = \langle (X-\langle X\rangle_R)(Y-\langle Y\rangle_R)\rangle_R\;.
\end{equation}

In this work, we are predominantly concerned with the computation of the chiral 
condensate, $\O=|\bar{\sigma}|$.
In \aref{app:complex_action}, we show exemplary data that we used to estimate the systematic
uncertainties arising from the second term in (\ref{eq:expectation_value_factorization}). 
In summary, we found that 
\begin{equation}
	\cov_R\left(e^{-\ii S_I},\vert\bar{\sigma}\vert\right)
  \sim \O\left(10^{-4}\dots10^{-2}\right)\;,
\end{equation}
while 
\begin{equation}
  \left\vert 1-\langle e^{-\ii S_I}\rangle_R\right\vert\sim\O\left(10^{-3}\dots10^{-1}\right)\;,
\end{equation}
such that overall we expect systematic uncertainties of
$\O\left(10^{-3}\right)$ from negligence of the complex-action problem, while our
statistical uncertainties are typically of $\O\left(10^{-2}\right)$. We conclude
that we may safely neglect the complex-action problem on our small and medium-sized
lattices. Future research on larger lattices might have to review this position, however.

	\section{Results}\label{sec:results}
	In the following, we present our lattice results for the chiral condensate in the parameter space 
spanned by the chemical potential $\mu$ and the magnetic field $B$ at a low temperature $T$. In 
particular, we aim at answering the question of what remains of the large\,-$\Nf$ phase structure 
shown in Figs. \ref{fig:pd_large_N} and \ref{fig:veff_large_N} when considering $\Nf=1$. Hence, we look 
for traces of inverse magnetic catalysis, multiple phase transitions in $\mu$ at $B\neq0$, and a 
first-order transition at $B=0$. Throughout, $\Ns$ and $\Nt$ denote the number of lattice points in 
each spatial and the temporal direction, respectively. Moreover, we employ periodic boundary 
conditions in space and anti-periodic ones in time for fermions, while the scalar field $\sigma$ is 
periodic in all directions.

We begin by showing in \fref{fig:cc_vs_mu_infVol} an infinite-volume extrapolation at fixed lattice 
spacing of $\sigex(\mu)$ for various values of $B$. In what follows, we shall discuss these results 
in more detail.

\subsection{Vanishing magnetic field}

\begin{figure}[t]
	\centering
	\includegraphics[width=\linewidth]{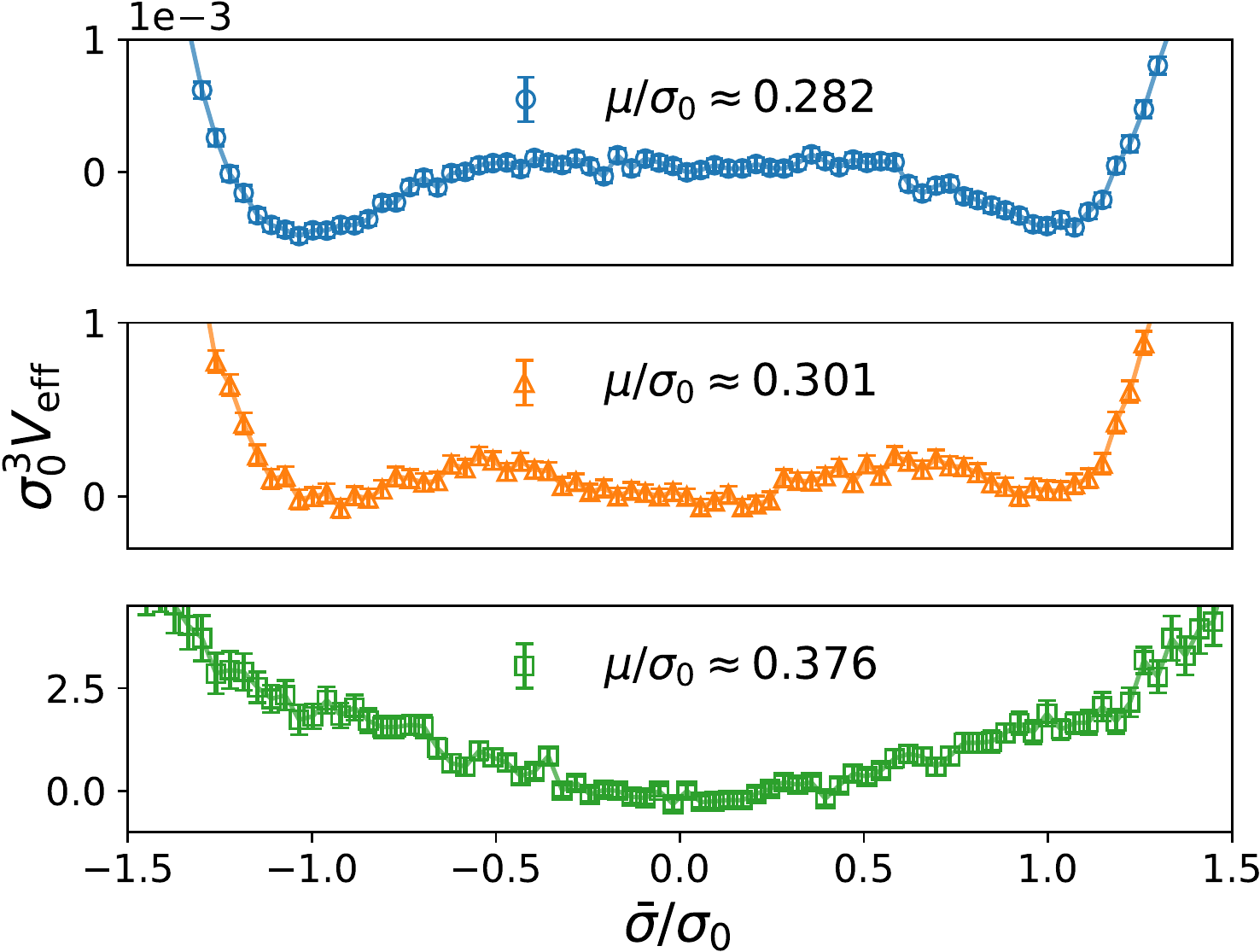}
	\caption{Constraint effective potential \cite{OWY86} for $B=0$, 
		determined as the logarithm of the probability 
	distribution of $\bar{\sigma}$ and normalized to zero at $\bar{\sigma}=0$; $\Ns=\Nt=8$ 
	and $a\sigma_0\approx1.063$\;.}
	\label{fig:veff_lattice}
\end{figure}

Focusing on $B=0$ first, one observes that, as anticipated, chiral symmetry is spontaneously broken 
at $\mu=0$, indicated by $\sigex\neq0$, and that the order parameter decreases with increasing $\mu$ 
(to a non-zero value due to our definition of $\sigex$). This behavior becomes sharper on larger 
volumes, which is the expected behavior for a phase transition. In order to determine the order of 
this transition, it is instructive to study histograms of $\bar{\sigma}$, as they allow one to 
reproduce the probability distribution $e^{-S_R}/Z_R$. We show the effective potential 
determined from this distribution in the vicinity of the phase transition on our 
smallest lattice in \fref{fig:veff_lattice}. The corresponding temperature amounts to 
$T/\sigma_0\approx0.118$; i.e., it is significantly different from zero.

\begin{figure*}[t]
	\subfloat{
		\includegraphics[width=0.35\linewidth]{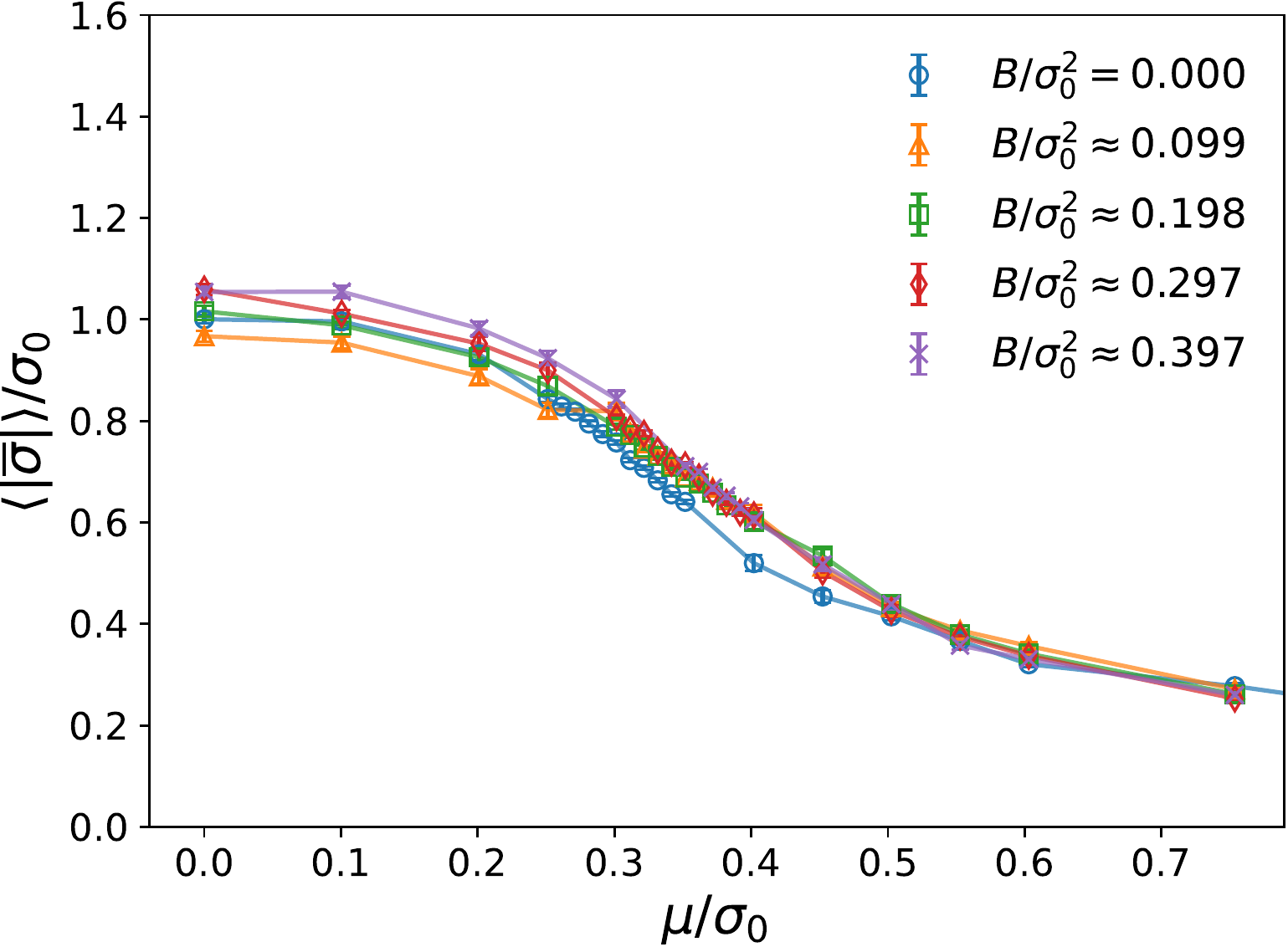}}
	\subfloat{
		\includegraphics[width=0.3105\linewidth]{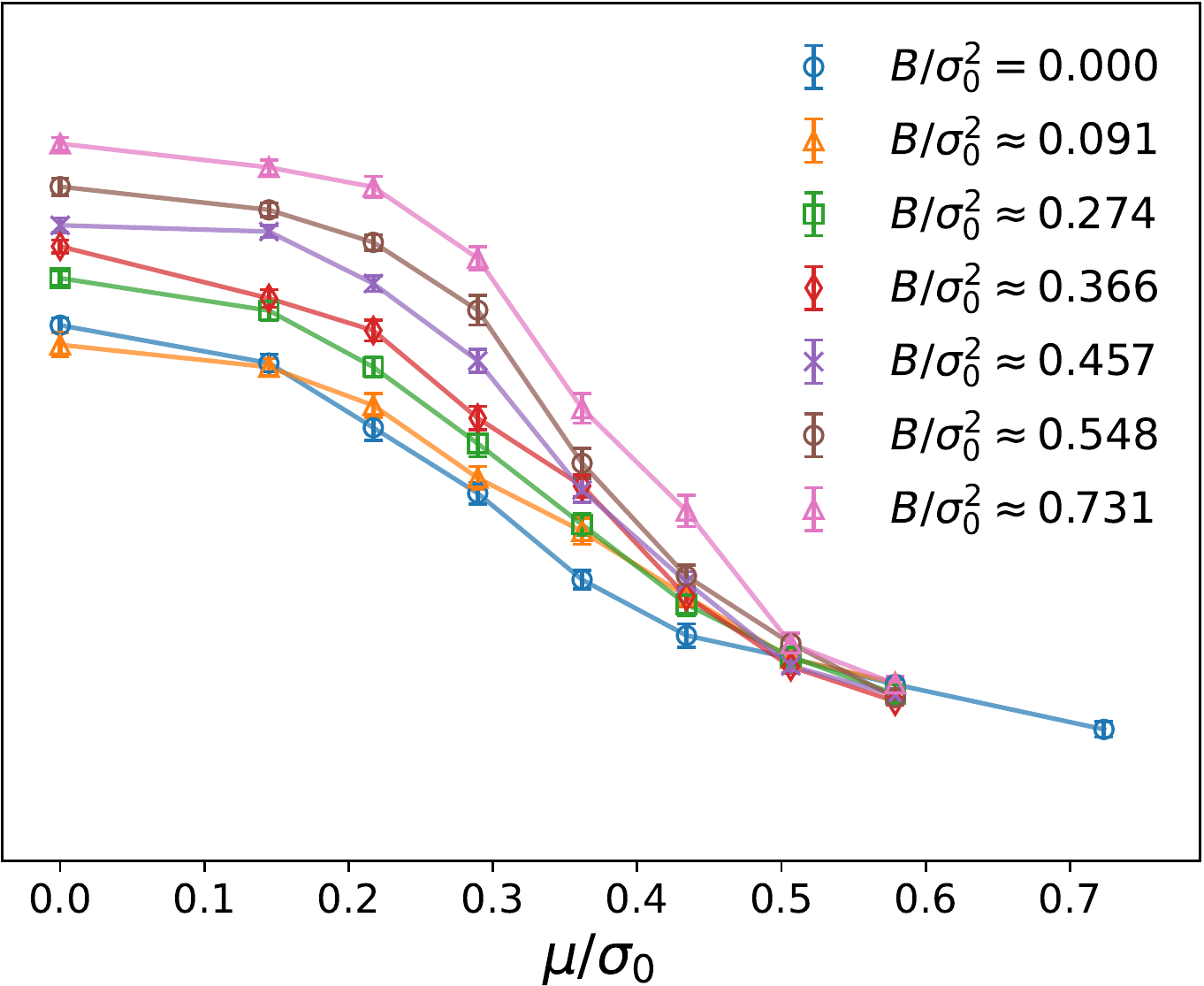}}
	\subfloat{
		\includegraphics[width=0.3105\linewidth]{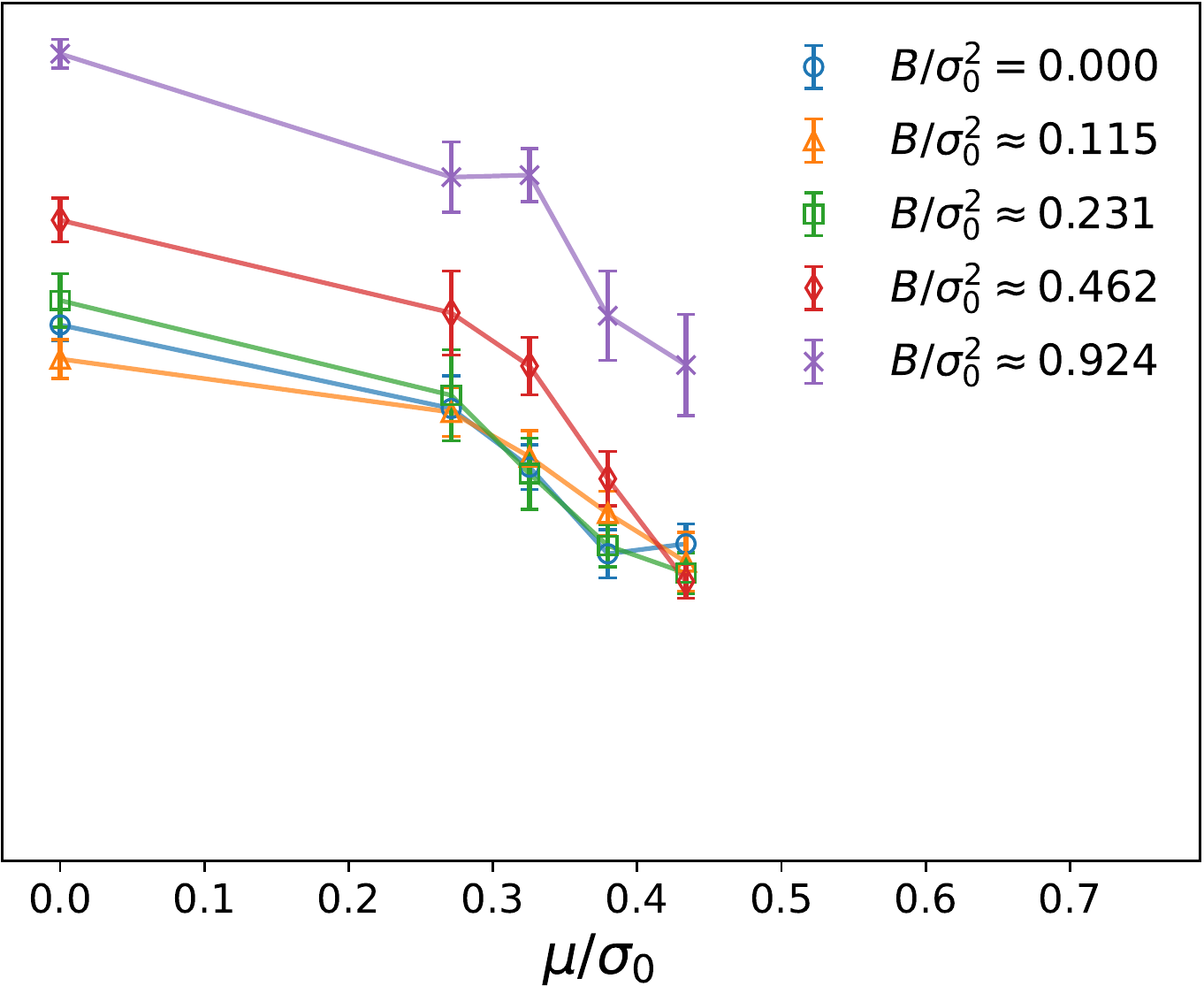}}
	\caption{Order parameter $\sigex$ as a function of $\mu$ for different magnetic field strengths 
	in a continuum extrapolation at fixed physical volume. (left) $\Ns=8$, $a\sigma_0\approx0.995$. (center) $\Ns=12$, $a\sigma_0\approx0.691$. (right) $\Ns=16$, $a\sigma_0\approx0.461$. We consider only cubic lattices 
	where $\Ns=\Nt$\;.}
	\label{fig:cc_vs_mu_cont}
\end{figure*}

One observes that for $\mu/\sigma_0\approx0.282$ the potential has two degenerate minima at 
$\bar{\sigma}/\sigma_0\approx\pm1$, while for $\mu/\sigma_0\approx0.301$ a third minimum emerges at 
$\bar{\sigma}=0$. Finally, at $\mu/\sigma_0\approx0.376$, only this trivial minimum is left. This 
is clear evidence for a first-order phase transition at $\mu=\muc\approx0.301\sigma_0$. That 
this is the case even on a lattice volume as small as $8^2$ does not really come as a surprise in 
light of the discussion regarding \fref{fig:veff_large_N} in \sref{sec:analytical}. We 
notice that the critical chemical potential $\muc$ is roughly $3$ times smaller than in the 
large\,-$\Nf$ limit. This reduction is less than the factor of around 4 or 5 found for the 
zero-density critical temperature in \cite{LMW23}.

We also remark that on small lattices the first-order transition is not visible from the 
$\mu$-dependence of $\sigex$ alone. However, as can be seen in \fref{fig:cc_vs_mu_infVol}, the transition becomes sharper for larger volumes and approaches the behavior one would expect from a first-order transition. This is due to the 
minima of $S_\mathrm{eff}$ deepening, making tunneling between them less probable. While a detailed 
analysis of the effective potential on our larger lattices was not possible due to limited 
statistics, we believe that the transition is of first order for all lattice sizes considered here.

Let us now compare our results at $B=0$ with the existing literature. Previous works using 
staggered fermions, \cite{HKK93} ($\Nf=12$) and \cite{KS01} ($\Nf=4$), report that the phase 
transition is of first order at $T\approx0$ (simulations at exactly vanishing temperature are, of 
course, impossible) and of second order for relatively high temperatures. They also claim that their 
findings are consistent with the existence of a first-order critical line at low temperatures, 
ending in a tri-critical point at some non-zero value of $T$, despite being unable to precisely 
locate this tri-critical point. These results were also confirmed by the OPT studies 
\cite{KPR07_2,KPR07}.

Lastly, we mention that our estimate of $\muc$, which is approximately $0.3\sigma_0$ for all 
lattice sizes considered, is significantly lower than the one quoted in \cite{HKK93}, where it is
comparable to the mean-field value, $\muc/\sigma_0=1$. However, this is expected since quantum 
fluctuations tend to destroy long-range order and, thus, shrink the region of broken chiral symmetry.

\subsection{Finite magnetic field}

Now turning to the case of $B\neq0$, we see from \fref{fig:cc_vs_mu_infVol} that for small $\mu$ the 
magnetic field has an overall tendency to increase the chiral condensate, corresponding to the 
magnetic catalysis scenario outlined in \sref{sec:analytical}. Notice, however, that on smaller 
lattices there are finite-volume effects that lead to a non-monotonic $B$-dependence of the chiral 
condensate: $\sigex$ first decreases for the weakest allowed non-vanishing magnetic field before 
increasing monotonically with $B$ for all stronger fields. This effect was also observed in our 
previous study \cite{LMW23} and can even be seen at the mean-field level but ceases to play a role 
for larger volumes. While this non-monotonicity could potentially be relevant for applications of 
the GN model in solid-state physics \cite{Liu99,HR08}, we do not discuss it further here as our 
work rather takes its motivation from high-energy physics, where one typically assumes infinite 
volumes.

Investigating larger chemical potentials next, the situation appears very much unchanged in that -- 
within the error margins -- the magnetic field increases the order parameter for all $\mu$ below 
the phase transition apart from possible finite-size effects. The transition itself appears to remain a 
(weak) first-order one even for $B\neq0$. Far beyond the phase transition, the magnetic field 
ceases to have a noticeable effect on $\sigex$, a behavior also observed in the finite\,-$T$ study 
\cite{LMW23}. We furthermore observe that the critical chemical potential of the transition 
slightly increases with $B$ within errors. 

These observations are in contradiction with the mean-field scenario of inverse magnetic catalysis 
discussed in \sref{sec:analytical}, as well as with the OPT study \cite{KPR13}, where inverse 
magnetic catalysis was predicted to persist even beyond the large\,-$\Nf$ limit. Furthermore, we 
find no evidence for multiple phase transitions in $\mu$ at finite $B$, whereas they were claimed 
in \cite{KPR13} to exist even for $\Nf=2$. While the situation could, in principle, be 
qualitatively different between $\Nf=2$ and $\Nf=1$, we consider this rather unlikely and are 
inclined to look for alternative explanations for this discrepancy.\footnote{There are ongoing
discussions about critical flavor numbers in related 2+1D Thirring models -- see, 
e.g., \cite{LWW19,WL22,Han19} -- but these are of a very different nature and we see no
indication for similar phenomena to arise in this context.}

It is important to stress that the missing features of inverse catalysis and
cascades of phase transitions are expected to happen in a very small parameter
	region.
In lack of a better guiding principle, one could assume that the
reduction of the critical chemical potential as a scale roughly carries over to
other features of the phase diagram. Even in the mean-field approximation they occur
only within $\pm10\dots20\%$ of the critical chemical potential -- being
themselves only a few percent; see \fref{fig:pd_large_N}. Scaling that down
leads us to expect filigree features of the size of $\mu/\sigma_0 \sim
\mathcal{O}(0.006)$ in a very small parameter regime of $\mathcal{O}(0.06)$ --
this is a scale that we cannot resolve with the current method assuming a
reasonable amount of resources. Even stronger physical constraints on the
sampling rate apply to the magnetic field's discretization due to the finite
volume. As the multiple phase transitions quickly oscillate in that direction of
the phase diagram, resolving, e.g., one of the spikes in the critical chemical
potential seems quite unlikely.

Nevertheless, we would have expected at least some kind of footprint and the
lack of any evidence begs an explanation. The main differences between
\cite{KPR13} and the present work are that the former was performed in the
continuum, in an infinite volume, and under the assumption that $\sigma$ is
homogeneous, while we consider the theory on a lattice of finite extent and
allow for arbitrary modulations of $\sigma$.  It is the latter difference, in
particular, that could be responsible for the absence of multiple phase
transitions, as strong fluctuations in $\sigma$ could likely wash out the
discrete Landau level structure, which is the origin of the cascade of transitions in the
first place. While we investigate the potential existence of inhomogeneities in
$\sigma$ below, we mention that the precise reason for the absence of inverse
magnetic catalysis and multiple phase transitions is still not entirely clear to
us. Additional work in that direction, however, is ongoing and intended to be
part of a forthcoming publication. It might also be enlightening to investigate
whether the OPT's prediction is stable with respect to the inclusion of higher
orders in $1/\Nf$ given the very small flavor numbers we are discussing here.

In order to study stronger magnetic fields (in units of $\sigma_0$), we also approach the continuum 
limit for fixed physical volume and show the results for the $\mu$-dependence of the order parameter 
in \fref{fig:cc_vs_mu_cont}. The observations are, however, in close analogy to the infinite-volume 
extrapolation discussed above: We find magnetic catalysis below the phase transition, a slight 
increase of $\muc$ with $B$, and no evidence for multiple phase transitions. 

\subsection{Inhomogeneous phases}
Lastly, as in \cite{LMW23}, we study the possibility of spatial inhomogeneities in $\sigma$ induced 
by the magnetic field. While the zero-density study \cite{LMW23} did not observe any evidence of 
such inhomogeneous structures, this is hardly surprising as even in the (four-dimensional) 
mean-field calculations \cite{FZK10,TNK15,BC16} inhomogeneities would arise only at $\mu\neq0$. 
Conceptionally speaking, the magnetic field is capable of inducing inhomogeneities due to the 
effective reduction of the number of space-time dimensions for strong enough $B$. This is because in 
low dimensions four-Fermi theories (at $B=0$) are known to develop inhomogeneous structures at 
finite density and low temperatures, as was first found in mean-field studies \cite{ST00,TU03} and 
has recently been confirmed by lattice simulations at finite $\Nf$ \cite{LPW20,LPW20_1,LMW22}. 

\begin{figure}[t]
	\includegraphics[width=0.95\linewidth]{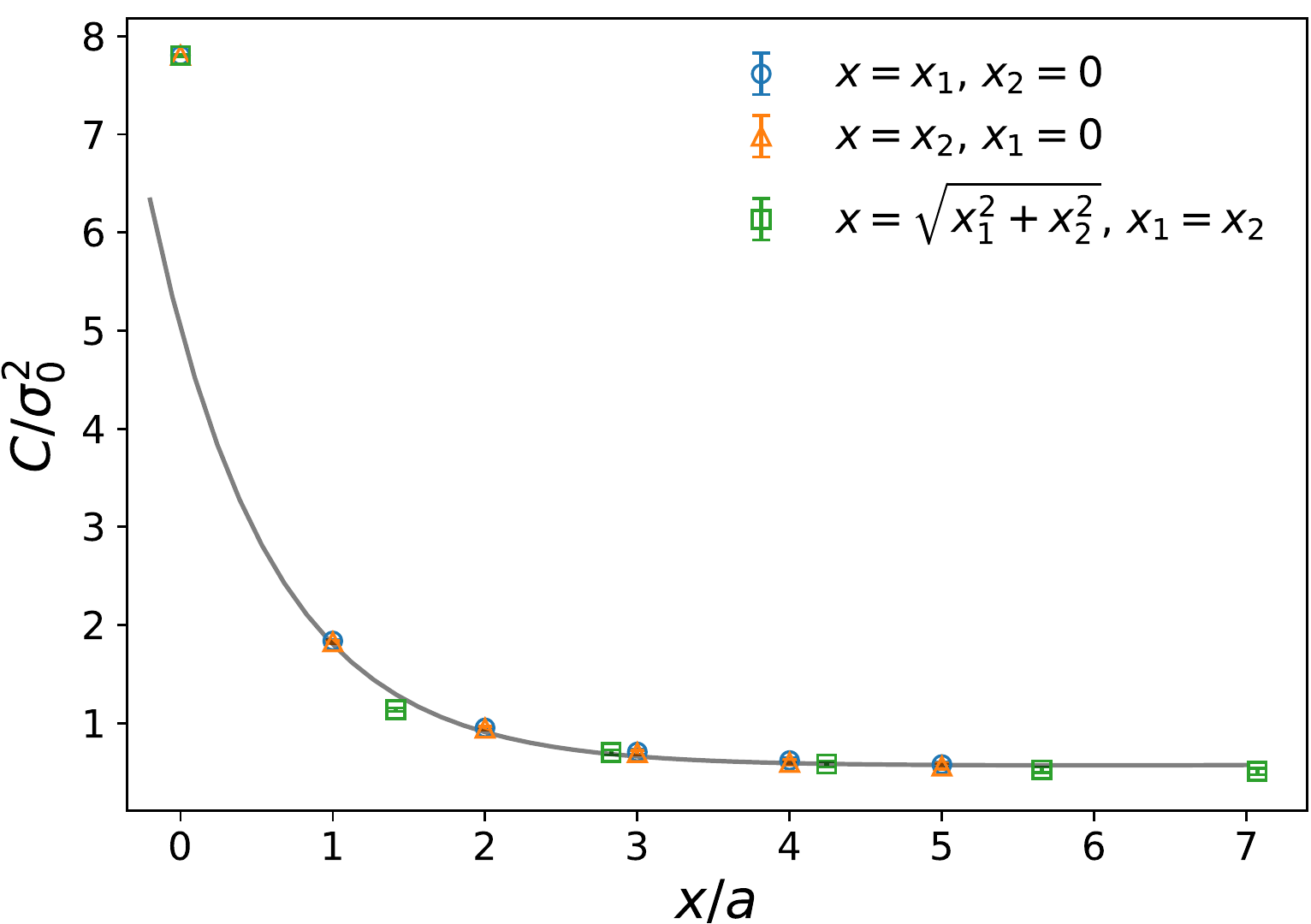}
	\caption{Spatial correlator $C$ from (\ref{eq:spatial_correlator}) along the coordinate axes 
	(either $x_1=0$ or $x_2=0$) as well as the diagonal ($x_1=x_2$) for $\Ns=\Nt=12$, $eB/\sigma_0^2\approx0.346$,
	$\mu/\sigma_0\approx0.349$, and $a\sigma_0\approx1.004$\;. The full line corresponds to a 
	$\cosh$ fit to the data points,	starting from $x\geq a$, and is included in order to guide the 
	eye and showcase rotational invariance.}
	\label{fig:spatial_correlator}
\end{figure}

In order to address this question, we follow \cite{LPW20,LMW23} in computing the spatial 
correlation function
\begin{align}\label{eq:spatial_correlator}
	\begin{aligned}
		C(x_1, &x_2) = \\
		\frac{1}{\Ns^2\Nt}&\sum_{x'}\left\langle\sigma(x'_0, x_1, x_2)\sigma(x'_0, 
		x_1+x'_1,x_2+x'_2)\right\rangle\;,
	\end{aligned}
\end{align}
where the $x_i$ denote the spatial components of a lattice point and the sum over $x'$ runs over the 
entire lattice. Any spatial inhomogeneities present in $\sigma$ should also be visible in $C$, but 
the latter has the advantage that it does not suffer from cancellations as the expectation value of 
$\sigma$ itself would \cite{LPW20}. However, as can be seen exemplarily in 
\fref{fig:spatial_correlator} for a relatively strong magnetic field at a low temperature and a 
chemical potential close to the phase transition we do not find any evidence for inhomogeneities 
even in our simulations at $\mu\neq0$. We have verified that the same is true for all of our other 
data points as well. We remark that earlier simulations of the model at vanishing magnetic field, 
using staggered fermions on larger lattices, did find signs of inhomogeneities \cite{HKS03}.

	\section{Discussion}\label{sec:discussion}
	In this work, we have presented the results of extensive lattice simulations of the Gross-Neveu model 
(\ref{eq:gn_model}) in $2+1$ dimensions at finite chemical potential and magnetic field, considering 
$\Nf=1$ flavor of four-component overlap fermions. We argue that the arising complex-action problem 
is under control. The $\mu\neq0$ simulations at $T\approx0$ discussed here complement our previous 
results \cite{LMW23} obtained at $\mu=0$ and $T\neq0$. 

It was the main goal of \cite{LMW23} and the present work to understand to what extent the rich 
phase structure in $(B,T,\mu)$ space the model exhibits in the mean-field limit ($\Nf\to\infty)$ 
persists when considering a finite flavor number. While at $\mu=0$ the generic features of the 
large\,-$\Nf$ model, i.e., magnetic catalysis for all temperatures below the phase transition and an 
increase of the critical temperature with $B$, were found to be exist also for $\Nf=1$ \cite{LMW23}, 
the situation at $\mu\neq0$ appears to be different. There, the mean-field approximation predicts a 
cascade of first-order phase transitions and a region of inverse magnetic catalysis, i.e., a 
decrease of the order parameter with $B$. However, we find no evidence of either of these effects in 
our simulations. On the contrary, we find magnetic catalysis below and in the vicinity of the phase 
transition and an increase of the critical chemical potential with $B$. While -- to the best of our 
knowledge -- no previous lattice results exist addressing this question, our findings are in 
contradiction with analytical results  using the OPT method to study the two-flavor theory, where 
the aforementioned features were found to persist \cite{KPR13}.

We have already mentioned the most substantial differences between \cite{KPR13} and the present work 
in that the former works in the continuum, in an infinite volume, and assumes translational 
invariance, $\sigma(x)=\sigma$, while we work on a lattice of finite extent and allow for $\sigma$ 
to vary in space and time. We believe that the latter difference could be responsible for the 
discrepancy, even though we find no trace of inhomogeneities at the level of Monte-Carlo averages. 
Still, random fluctuations in $\sigma$ could conceivably be strong enough to completely smear out 
the discretized energy levels induced by the Landau quantization causing the multiple-transition 
pattern. Moreover, the argument explaining the origin of inverse magnetic catalysis in \cite{PRS13} 
was also given under the assumption that the order parameter is constant, which could potentially
invalidate it in the context of our lattice studies. 

Another, perhaps less interesting, explanation is that the OPT method might plainly be not 
reliable for such small
flavor numbers anymore. A follow-up study dealing with this discrepancy with 
more rigor, also considering larger $\Nf$, is currently ongoing. We mention in passing that our 
simulations were performed within the strong-coupling regime, where there is spontaneous symmetry 
breaking for $T=0$ and $\mu=0$ even at vanishing magnetic field. However, our observation that 
inverse magnetic catalysis and the cascade of phase transitions are both absent rather resembles the 
situation in the weak-coupling regime, where chiral symmetry is intact when $T$, $\mu$ and $B$ all 
vanish and is broken only for non-zero magnetic field \cite{KPR13}.

We found that the phase transition in $\mu$ for all magnetic field strengths likely is of weak first order even at 
relatively high temperatures, $T/\sigma_0\approx0.1$, which in the mean-field limit happens only
for non-vanishing $B$. Our results are, however, in agreement with the previous $B=0$ lattice 
\cite{HKK93,KS01} and OPT \cite{KPR07_2,KPR07} studies, the latter of which claim that this 
first-order phase transition is a consequence of $1/\Nf$ corrections. However, we argue that on a 
finite volume a first-order transition at $B=0$ and $T\neq0$ can emerge in the large\,-$\Nf$ limit 
as well.

Lastly, we briefly comment on the potential relevance of our results for QCD. The results of 
\cite{LMW23}, predicting magnetic catalysis for every temperature below the phase transition, are in 
disagreement with the inverse magnetic catalysis scenario taking place around the chiral crossover 
in QCD \cite{BBE12_2,BBE12}. This discrepancy can be understood by the fact that around the 
crossover the ``sea quark'' contribution, encoding the back-reaction of the (charged) quarks onto 
the (neutral) gluonic distribution, dominates over the ``valence quark'' contribution, which causes 
an enhancement of chiral symmetry breaking \cite{BEK13}. This dominance of the sea quark effect then 
causes the chiral condensate to decrease. On the other hand, the purely fermionic GN model is, 
without modifications (see, e.g., \cite{AHL21,And21r,BF21}), obviously incapable of reproducing this 
gluon-induced phenomenology. 

For temperatures far above and below the QCD crossover, however, the valence contribution 
dominates. If this were true at finite density as well, one could speculate that gluonic effects 
might be less relevant for the low-temperature regime at finite chemical potential we studied in 
this work. This would imply that our results might be of relevance for the finite-density regime of 
QCD, at least on a qualitative level. However, further research in that direction is certainly 
necessary in order to make any definite statements, since our $(2+1)$-dimensional model at $\Nf=1$ 
is clearly still quite different from QCD.
	
	\begin{acknowledgments}
		This work would not have been possible without the simulation framework provided by 		
		Björn Wellegehausen. M. M. is indebted to Laurin Pannullo, Malte Schulze, Ivan Soler, and 
		Marc Winstel for useful discussions. J. J. L. thanks Ed Bennett for helpful discussions about
the reproducibility and openness of this publication.

		This work has been funded by the Deutsche Forschungsgemeinschaft (DFG) under 
		Grant No. 406116891 within the Research Training Group RTG 2522/1. 
    The work of J. J. L. was supported by the UKRI Science and Technology
    Facilities Council (STFC) Research Software Engineering Fellowship
    EP/V052489/1 and by the Supercomputing Wales project, which is part-funded
    by the European Regional Development Fund (ERDF) via Welsh Government.
		The simulations were performed on resources of the Friedrich Schiller 
		University in Jena supported in part by the DFG Grants No. INST 275/334-1 FUGG and No.
    INST 275/363-1 FUGG, as well as on the Swansea University SUNBIRD cluster
    (part of the Supercomputing Wales project). The Swansea University SUNBIRD
    system is part funded by the European Regional Development Fund (ERDF) via
    Welsh Government.
	\end{acknowledgments}
	
	\section*{Open Access Statement}
  		For the purpose of open access, the authors have applied a Creative Commons Attribution 
  		(CC BY) license to any author accepted manuscript version arising.

  	\section*{Data Availability Statement}
  		Full data underlying this work are available at Ref.~\cite{data}. Fully
  		automated analysis workflows can be found at Ref.~\cite{code}. Raw data and
  		the simulation code for generating the configurations are available upon
  		request.	
	
	\appendix
	\section{Complex-action problem}\label{app:complex_action}
	In the main text, we hinted that, while there is a non-negligible complex-action
problem of up to $10\%$, the complex phase of the action is almost uncorrelated
with the chiral condensate $\sigex$, the observable of predominant interest in this
paper. In this scenario, it would be justifiable to
neglect the complex-action problem for estimation of this particular observable,
and we shall establish in this appendix that this is indeed the case.

Because of the significant computational cost, we have estimated the complex phase
for a given ensemble only on a randomly drawn subset of configurations. We have
always made sure that the number of randomly drawn configurations is
significantly smaller than the effective number of statistically independent
configurations while being large enough for reliable statistical estimates. For
details of this procedure, we refer the reader to the corresponding code
publication \cite{code}. In the following, all analysis is done with respect to such
sub-ensembles.

\begin{figure}[t]
	\centering
	 \includegraphics[width=\linewidth]{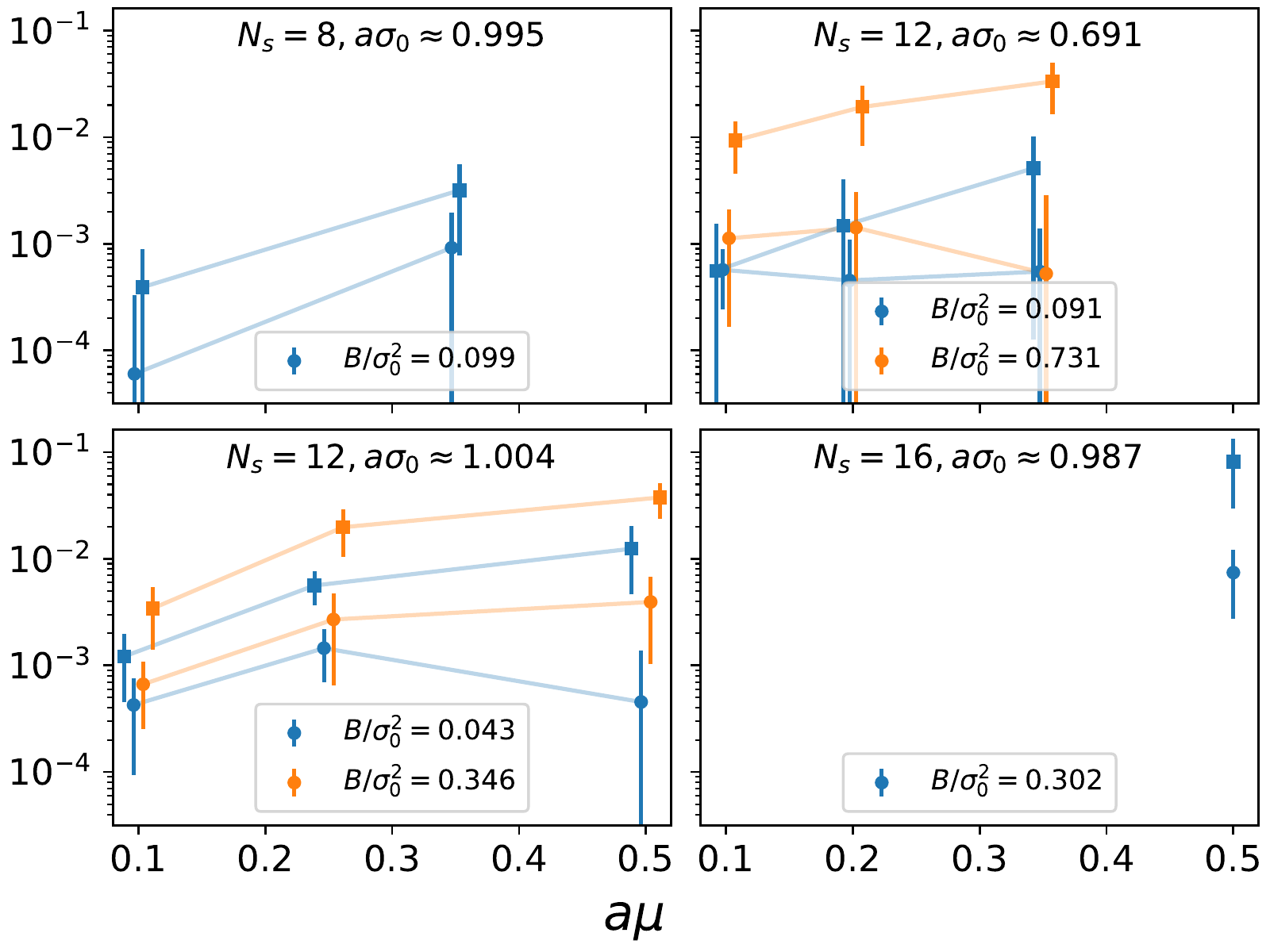}
    \caption{Average phase (squares) and covariance between chiral condensate and complex
  	phase of the action (circles) for different lattices over the chemical potential. Points
  	are horizontally shifted for better visibility.}
  \label{fig:complex_action_problem}
\end{figure}

\fref{fig:complex_action_problem} shows the average phase and the covariance 
(\ref{eq:covariance})
between $|\bar{\sigma}|$ and the complex phase of the action. The four panels
show different lattices. Because of the excessive numerical cost, we could provide only a single data point
for $\Ns=16$ for consistency checks. One should note that we have taken the
absolute value here in order to capture the maximal possible effect in a single
(real) number. Both quantities are complex valued in general.

The data show the expected trend: The average phase becomes smaller (i.e., the
complex-action problem becomes more severe) with growing $\mu$ and/or $B$. The maximal
deviation of the average phase from unity is roughly $0.1$ for the largest $\mu$ and
$B$ considered. 

The covariance consistently is about an order of magnitude smaller on average and
often even compatible with 0. One could conjecture that there is an analytical
argument for the independence of the average phase and $\sigex$, but we have 
not found one yet. Moreover, the errors of both quantities are well under
control, although the sample size for $\Ns=16$ is very small due to the
excessive cost of computing the full determinant.

	\section{Parameters}\label{app:parameters}
	In \tref{tab:parameters}, we list the parameters used in our simulations. The magnetic flux quantum 
number $b$ appears in the quantization condition of the magnetic field on the lattice:
\begin{equation}\label{eq:magnetic_field_quantization}
	eB = \frac{2\pi}{L^2}b\;,
\end{equation}
with $L^2$ denoting the area of the spatial plane.
For further details on the simulations and the 
way we perform our error analysis, we refer to \cite{LMW23}.

\begin{table*}
	\caption{Parameters we have generated configurations for. $\Ns$ ($\Nt$) denotes the number of 
	lattice points in spatial (temporal) direction, $g^2$ is the four-Fermi coupling in 
	\eqref{eq:gn_model}, $b$ denotes the magnetic flux quantum number in \eqref{eq:magnetic_field_quantization}, $a\mu$ is the chemical 
	potential in lattice units, and $T_0$ denotes the temperature at which we set the scale 
	$\sigma_0$, which we give in lattice units in the last column. Notice that $T_0$ differs between 
	the infinite-volume and continuum extrapolations; see \cite{LMW23}.}
	\renewcommand{\arraystretch}{1.5}
	\renewcommand{\tabcolsep}{8pt}
	\newcommand{\ncol}{7}
	\begin{tabular}{ccccccc}
		\hline\hline
		$\Ns$ & $\Nt$ & $1/g^2$ & $b$ & $a\mu$ & $T_0/\sigma_0$ & $a\sigma_0$ \\
		\hline
		\multicolumn{\ncol}{c}{\textbf{infinite-volume extrapolation}} \\
		\hline\\[-0.5cm]
		$8$ & $8$ & $0.1520$ & \makecell{$0$, $1$, $2$, $3$, $4$ \\ $0$ \\ $1$, $2$ \\ $3$ \\ $4$} 
		& 
		\makecell{$0.00$, $0.10$, $0.20$, $0.25$, $\hdots\,$, $0.60$, $0.75$ \\ $0.26$, $0.27$, $\hdots\,$, $0.34$ \\ $0.31$, $0.32$, $\hdots\,$, $0.38$ \\ $0.31$, $0.32$, $\hdots\,$, $0.39$ \\ $0.36$, $0.37$, $0.38$, $0.39$} & $0.059$ & $1.063$ \\
		$8$ & $16$ & $0.1520$ & $0$ & $0$ & $0.059$ & $1.063$ \\
		\hline\\[-0.5cm]
		$12$ & $12$ & $0.1520$ & 
		\makecell{$0$, $1$, $2$, $3$, $4$, $6$, $8$ \\ $0$, $1$, $3$, $6$, $8$ \\ $6$, $8$} &
		\makecell{$0.00$ \\ $0.10$, $0.20$, $0.25$, $\hdots\,$, $0.40$, $0.50$, $0.60$, $0.75$ \\ $0.33$, $0.34$, $\hdots\,$, $0.42$} &
		$0.062$ & $1.004$\\
		$12$ & $16$ & $0.1520$ & $0$ & $0$ & $0.062$ & $1.004$ \\
		\hline
		$16$ & $16$ & $0.1520$ & 
		\makecell{$0$, $1$, $2$, $4$, $8$, $12$, $16$ \\ $0$, $1$, $2$, $4$, $8$, $12$} &
		\makecell{$0.00$ \\ $0.25$, $0.30$, $0.40$, $0.50$} &	$0.064$ & $0.984$\\
		\hline
		\multicolumn{\ncol}{c}{\textbf{continuum extrapolation}}\\
		\hline
		$8$ & $8$ & $0.1520$ & \multicolumn{2}{c}{(see above)} & $0.126$ & 
		$0.995$\\
		\hline\\[-0.5cm]
		$12$ & $12$ & $0.1650$ & 
		\makecell{$0$ \\ $1$, $3$, $4$, $5$, $6$, $8$ \\ $2$} &
		\makecell{$0.00$, $0.10$, $0.15$, $\hdots\,$, $0.40$, $0.50$ \\ $0.00$, $0.10$, $0.15$, $\hdots\,$, $0.40$ \\ $0.00$} & $0.121$ & $0.691$\\
		\hline\\[-0.5cm]
		$16$ & $16$ & $0.1740$ & 
		\makecell{$0$, $1$, $2$, $3$, $4$, $6$, $8$, $12$, $14$, $16$ \\ $0$, $1$, $2$, $4$, $8$} &
		\makecell{$0.00$ \\ $0.125$, $0.150$, $0.175$, $0.200$} & $0.136$ & $0.461$\\
		\hline
		\hline
	\end{tabular}
	\label{tab:parameters}
\end{table*}

	%\onecolumngrid
	%\FloatBarrier	
	%\twocolumngrid	

	\bibliography{bibliography}

\end{document}